\documentclass[journal,final,twoside]{IEEEtranTCOM}
\normalsize

\usepackage{cite}

\usepackage[pdftex]{graphicx}
\graphicspath{{Figures/}}
\DeclareGraphicsExtensions{.pdf}

\usepackage[cmex10]{amsmath}

\usepackage{amsthm}
\usepackage{amsfonts, amssymb, mathtools}
\usepackage{cases}
\usepackage{bm, bbm}

\usepackage{algorithm}
\usepackage{algorithmic}

\usepackage{array}
\usepackage{multirow}

\usepackage[font=footnotesize]{subfig}

\usepackage{stfloats}

\usepackage{url}

\newcommand{\mbs}[1]{\pmb{#1}}
\newcommand{\vect}[1]{{\lowercase{\mbs{#1}}}}

\newcommand{\1}{\mathbbm{1}}
\newcommand{\rv}[1]{{\mathrm{#1}}}

\newcommand{\av}{\vect{a}}
\newcommand{\hv}{\vect{h}}
\newcommand{\xv}{\vect{x}}
\newcommand{\thetav}{\vect{\theta}}

\newcommand{\abss}{\text{abs}}
\newcommand{\ang}{\text{ang}}

\newcommand{\MSCD}{\mathrm{MSCD}}

\newcommand{\Nt}{N_{\text{t}}}
\newcommand{\Nr}{N_{\text{r}}}

\newtheorem{lemma}{Lemma}

\theoremstyle{definition}
\newtheorem{definition}{Definition}

\newtheorem{remark}{Remark}

\newcounter{tempeqcounter}

\begin{document}

\title{Context-Tree-Based Lossy Compression and\\ Its Application to CSI Representation}

\author{Henrique~K.~Miyamoto,~\IEEEmembership{Graduate Student Member,~IEEE,}
        and~Sheng~Yang,~\IEEEmembership{Member,~IEEE}%
        \thanks{An earlier version of this paper was presented in part at the International Zurich Seminar on Information and Communication (IZS 2022)~\cite{miyamoto}.}
		\thanks{H.~K.~Miyamoto was with with the Laboratory of Signals and Systems~(L2S), CentraleSup\'elec, Paris-Saclay University, 91190 Gif-sur-Yvette, France. He is now with the Institute of Mathematics, Statistics and Scientific Computing~(IMECC), University of Campinas (Unicamp), Campinas, 13083-859, Brazil (email: hmiyamoto@ime.unicamp.br).}%
		\thanks{S.~Yang is with the Laboratory of Signals and Systems (L2S), CentraleSup\'elec, Paris-Saclay University, 91190 Gif-sur-Yvette, France (email: sheng.yang@centralesupelec.fr).}%
		\thanks{\textcopyright~2022 IEEE. Personal use of this material is permitted. However, permission to use this material for any other purposes must be obtained from the IEEE by sending a request to pubs-permissions@ieee.org.}%
}

\maketitle

\begin{abstract}
	We propose novel compression algorithms for time-varying channel state information~(CSI) in wireless communications. The proposed scheme combines (lossy)~vector quantisation and (lossless)~compression. First, the new vector quantisation technique is based on a class of parametrised companders applied on each component of the normalised CSI vector. Our algorithm chooses a suitable compander in an intuitively simple way whenever empirical data are available. Then, the sequences of quantisation indices are compressed using a context-tree-based approach. Essentially, we update the estimate of the conditional distribution of the source at each instant and encode the current symbol with the estimated distribution. The algorithms have low complexity, are linear-time in both the spatial dimension and time duration, and can be implemented in an online fashion. We run simulations to demonstrate the effectiveness of the proposed algorithms in such scenarios.
\end{abstract}


\section{Introduction}

\IEEEPARstart{W}{ireless} communication systems feature an ever growing dimension due to larger antenna arrays, denser network deployments, and an increasing number of terminals and devices. To maintain connectivity in such systems, a colossal amount of channel measurements, often referred to as channel state information~(CSI), are necessary. Efficiently representing the CSI is crucial for storage and dissemination. A typical example is the downlink transmission from a base station~(BS) with multiple antennas~(potentially a large number, i.e., massive MIMO) to multiple users simultaneously. The BS should steer the signal for user~$j$ in such a way that the interference with any other user $k\ne j$ is low enough. Such beamforming techniques rely on precise CSI at the transmitter side, e.g.,~\cite{caire}. For the BS to acquire the CSI, however, it usually requires that each user feeds back the CSI measurements in a timely and accurate fashion. How to reduce the bandwidth cost of such feedback traffic, which is highly non-negligible, is becoming a crucial problem. This is essentially a lossy data compression problem.

CSI measurements are typically correlated in space and time according to the propagation environment, and the mobility of users and obstacles. The spatial correlation for a single antenna array is inherent to the antenna structure and can be used to reduce CSI dimension so that only a few coefficients are needed to describe the channel state. For large antenna arrays, recent works apply deep learning and compressed sensing techniques to further exploit the correlation and sparsity of channel, e.g.,~\cite{wen, guo, mashhadi} and references therein. Independent channel coefficients~(e.g., when correlation is ignored or after decorrelation) are then quantised with a vector quantiser into symbols from a finite set~(codebook), e.g.,~\cite{decurninge, shlezinger} and references therein. Further spatial compression can be achieved with entropic encoding~(e.g., arithmetic coding) on the bit representation of the quantisation indices~\cite{mashhadi}.

The temporal correlation of CSI measurements, on the other hand, is less exploited for feedback. Indeed, the sequence of quantised symbols, considered as a random process, can be losslessly compressed to a bit-stream. If the sequence is stationary, then the bit rate can theoretically be as low as the entropy rate of the underlying process. A possible approach towards CSI compression is therefore to directly apply any universal compression algorithm~\cite{gassiat, cover, csiszar}, such as Lempel-Ziv~\cite{ziv-77, ziv-78}~(known as LZ77 and LZ78) to the quantisation indices.

Another universal compressor is the \emph{context-tree weighting}~(CTW) algorithm~\cite{willems-ctw}, which learns the distribution of a given sequence in an efficient way. The learned distribution can then be used as the coding distribution to compress the sequence in combination with arithmetic coding. It has been shown that, in this case, Rissanen lower bound~\cite{willems-ctw} is achieved, in the sense of having optimal rate of convergence to the entropy for tree sources with unknown parameters. Extensions of the algorithm can be found in~\cite{tjalkens-multi1, tjalkens-multi2, tjalkens-multi3, begleiter1, begleiter2}. A modification of CTW based on the minimum-description principle yields the \emph{context-tree maximising}~(CTM) algorithm~\cite{willems-ctm}, which can produce maximum \textit{a posteriori}~(MAP) probability tree models~\cite{willems-map}. Connections between CTW/CTM algorithms and Bayesian inference have been explored in~\cite{gassiat, mertzanis, kontoyiannis}. In particular, in~\cite{kontoyiannis}, the authors extended the CTM algorithm to find the $k$ \emph{a posteriori} most likely models, under the name of \emph{Bayesian context-tree}, and generalised some results.

Directly applying these algorithms to compress quantisation indices presents, nonetheless, some difficulties. First, the output bit-stream is of variable length, making the feedback difficult to implement. Second, in Lempel-Ziv methods, the input symbol block is also of variable length, since it depends on parsing the original sequence. This means that the encoder may need to wait for an indefinite number of time slots to output an indefinite number of bits for feedback. Finally, arithmetic coding assumes that computations are carried out with infinite precision, while, in practice, it has to be carefully implemented so as to deal with finite precision constraints, e.g.,~\cite{witten, moffat}. Trying to avoid such difficulties motivates us to propose new compression algorithms adapted to communication scenarios.

In this work, we focus on the problem of online lossy compression of a sequence of CSI vectors, for which we propose a two-step solution. The first step is lossy: we normalise the CSI vector and quantise the amplitude and phase components separately using a data-adapted compander, followed by uniform quantiser. In particular, we consider the widely used $\mu$-compander and a new one called $\beta$-compander, inspired by the beta distribution. The second step is lossless: we compress the sequences of quantisation indices with the coding distribution estimated via a context-tree method. Two solutions can be considered: 1) to directly use CTW with arithmetic coding, or 2) to apply CTM to estimate the conditional distribution of the upcoming symbol at each time instant, and use this probability to compress the symbol. In the second case, we encode each symbol with a fixed number of levels to limit the fluctuation of the encoded bits flow, which is a desirable property in communication systems.

An important difference from previous works using deep learning techniques~\cite{wen, guo, mashhadi} is that they consider almost static channels, whereas our work investigates low, medium and high mobility scenarios. In addition, our scheme requires much less training samples: in fact, it can even be initialised without any training data, and training can be done online, as the sequence of coefficients is observed.

Our algorithms are linear-time in both the spatial dimension and time duration, and can be implemented in an online fashion. Although we propose the two steps as an ensemble, they are actually modular. This means that the new quantiser design and the new compression algorithms can be used independently, and combined with other existing quantisation or compression methods, if desired. Implementation codes are available in~\cite{webpage}.

The remainder of the paper is organised as follows. In Section~\ref{sec:preliminaries} we introduce the system model and review basic concepts of vector quantisation and context-tree representation. Our quantiser design is described in Section~\ref{sec:quantisation}, while the compression algorithm is presented in Section~\ref{sec:compressor}. Numerical simulations of CSI acquisition are analysed in Section~\ref{sec:results}. Finally, we draw some conclusions in Section~\ref{sec:conclusion}.

\subsubsection*{Notation}

Throughout this paper, we use the following notational conventions. Vectors are denoted by bold italic lower-case~(e.g., $\pmb{v}$), and their $L_2$-norm is denoted by $\| \pmb{v} \|$. Random variables are denoted by non-italic upper case letters~(e.g., $\rv{X}$). A binary string is denoted by bold non-italic lower case~(e.g, $\mathbf{c}$). Logarithms are to the base $2$. We denote $[n] \coloneqq \{1,\ldots,n\}$. The indicator function $\1_{\{P\}}$ takes value $1$ if the argument $P$ is a true statement, and $0$ otherwise.

\section{Problem Formulation and Preliminaries} \label{sec:preliminaries}

\subsection{Main Problem} \label{subsec:model}

Let us consider a network composed of a transmitter~(e.g., base station) and $\Nr$ receivers (e.g., mobile users). Assume that the \emph{channel state information}~(CSI) between the transmitter and receiver~$k$ at time~$t$ can be described by a complex vector $\hv_k[t]\in\mathbb{C}^{\Nt \times 1}$, for $k\in[\Nr]$. The dimension $\Nt$ of the vector can depend on the number of antennas and subcarriers used by the transmitter. For different purposes, such as feedback and storage, each receiver is required to represent its state sequence in an `economical' way, i.e., to use as few bits as possible to describe the sequence, for a given distortion constraint. This is known as the lossy source coding problem, and the fundamental trade-off between the rate of the encoded sequence and the distortion for a stationary process is known for a given distribution~\cite{cover}.

In this work, we are interested in compression algorithms for a source with unknown distribution that can be implemented in an online fashion, i.e., each $\hv_k[t]$ can be successively compressed, while exploiting the time correlation of the sequence. In most practical scenarios, the norm of the vectors $\hv_k$~(i.e., the strength) is less important than the relative strength of the components~(i.e., the direction). Therefore, our goal is to compress the normalised vector $\hv_k[t]/\|\hv_k[t]\|$. Before presenting our scheme, we recall some basic notions of vector quantisation, lossless compression and context-tree representation.

\subsection{Vector Quantisation} 

A \textit{vector quantiser}~\cite{gersho} of dimension $p$ and size $M$, is a mapping $q: \mathbb{R}^p \to \mathcal{C} \coloneqq \{ \pmb{y}_0, \pmb{y}_1, \dots, \pmb{y}_{M-1} \} \subset \mathbb{R}^p$ that assigns each vector $\pmb{x} \in \mathbb{R}^p$ to a codeword $\hat{\pmb{x}} \coloneqq q(\pmb{x}) = \pmb{y}_k$, for some $k \in \{0, 1, \dots, M-1\}$. To a sequence of vector symbols ${\pmb{x}_1^n \coloneqq \pmb{x}_1 \pmb{x}_2 \cdots \pmb{x}_n}$ we can apply vector-by-vector quantisation. In this case, the vector quantiser outputs a sequence of quantised vectors ${\hat{\pmb{x}}_1^n \coloneqq \hat{\pmb{x}}_1 \hat{\pmb{x}}_2 \cdots \hat{\pmb{x}}_n}$ and a sequence of quantisation indices ${k_1^n \coloneqq k_1 k_2 \cdots k_n}$, where $\hat{\pmb{x}}_i = \pmb{y}_{k_i}$, for each $i \in [n]$.

Two important parameters to assess the performance of a vector quantiser are the quantisation rate and the mean distortion. The \textit{quantisation rate}, defined as $R \coloneqq (\log_2 M)/p$, is an indicator of the cost to describe the vector, while the \textit{mean distortion} measures the error induced by the quantisation. A commonly used distortion measure is the mean squared chordal distance~(MSCD). Specifically, the MSCD between the original vector $\xv$ and the quantised vector $\hat{\xv}$ is defined as
\begin{equation}
\MSCD({\xv}, \hat{\xv}) \coloneqq 1 - \mathbb{E} \left[ \frac{{|\langle {\xv}, \hat{\xv} \rangle|}^{2}}{\|\xv\|^2 \|\hat{\xv}\|^2}  \right].
\end{equation}%
Note that the MSCD is invariant to scalar rotations, what is adapted to our application of CSI representation.

\subsection{Lossless Compression and Universality}

Consider a sequence of symbols (e.g., quantisation indices) from an $m$-ary discrete alphabet $\mathcal{A}=\{0,\ldots,m-1\}$. It is well known that, in the context of source coding, if the distribution $P$ of a source is known, Shannon's code can be used to generate a codeword with length $\lceil - \log P({x}_1^n) \rceil$ for any source sequence~$x_1^n$, where $P(x_1^n)$ is the probability of the realisation $x_1^n$. The expected length of such a code is within $1$~bit of the entropy lower bound $H(\rv{X}_1^n) \coloneqq \mathbb{E}[- \log P(\rv{X}^n_1)]$. Therefore, the coding rate, as the number of encoded bits per input symbol, can be arbitrarily close to $\frac{1}{n}H(\rv{X}_1^n)$. 

If, however, $P$ is not known and another probability distribution~(also called coding distribution) $Q_n$ is used instead, the codeword length becomes $\lceil - \log Q_n({x}_1^n) \rceil$, incurring a \emph{redundancy}
\begin{align}
	R(P, Q_n) &\coloneqq \mathbb{E}_P[-\log Q_n(\rv{X}_{1}^{n})] -	\mathbb{E}_P[-\log P( \rv{X}_{1}^{n} ) ] \nonumber\\
	&= D(P \, \| \, Q_n),
\end{align}
which coincides with the Kullback-Leibler divergence $D(P \, \| \, Q_n)$ between $P$ and $Q_n$. Without the knowledge of $P$, it is desirable to have low redundancy for every distribution in a given class of distributions. A coding distribution $Q_n$~(and the corresponding code) is said to be \emph{(weakly) universal}~\cite{csiszar} for a class $\mathcal{P}$ of processes if $\frac{1}{n} R(P, Q_n) \to 0, \ \forall P \in \mathcal{P}$. For instance, both the Lempel-Ziv codes and the CTW algorithm with arithmetic coding are universal for the class of stationary ergodic sources~\cite{gassiat, cover, willems-ctw}.

\subsection{Variable-Order Markov Chain and Context-Tree Representation}
\label{subsection: markov-chain}

Let us denote ${x}_{i}^{j} \coloneqq x_i x_{i+1} \cdots x_{j}$ a scalar sequence over an $m$-ary alphabet ${\mathcal{A} = \{0, 1, \dots, m-1\}}$, generated by a source with probability distribution $P$. We denote $l({x}_{i}^{j}) \coloneqq j-i+1$ the length of sequence ${x}_{i}^{j}$. A \textit{variable-order Markov chain} with order or memory $D$~(also called \textit{bounded memory tree source}) is a random process for which the probability of a new symbol, given the whole past, only depends on the last $D$ symbols, i.e., $P(x_{i}|{x}_{-\infty}^{i-1}) = P(x_{i}|{x}_{i-D}^{i-1})$. The main reason for our interest in Markov chains here is that any stationary ergodic source can be approximated by a Markov chain with sufficiently large order. Specifically, the entropy rate of a $D\text{-th}$ order Markov chain approximation of a stationary ergodic process becomes arbitrarily close to that of the original process when ${D\to\infty}$~\cite{gassiat, cover}. In many practical cases, a small $D$ is enough to describe a given process. 

The statistical behaviour of a variable-order Markov chain can be described by a \textit{context set}~$\mathcal{S}$~(also known as \textit{suffix set} or \textit{model}), which is a subset of $\bigcup_{i=0}^{D} \mathcal{A}^{i}$ that is proper~(no element in $\mathcal{S}$ is a proper suffix of any other) and complete~(each ${x}_{-\infty}^{n}$ has a suffix in $\mathcal{S}$, which is unique by properness). The \textit{context function} $c : \mathcal{A}^D \rightarrow \mathcal{S}$ maps each context ${x}_{i-D}^{i-1}$ with length $D$ to a suffix $c({x}_{-\infty}^{i-1}) = c({x}_{i-D}^{i-1}) = {x}_{i-j}^{i-1},\ j \le D$.
Furthermore, each suffix ${s} \in \mathcal{S}$ is associated with a parameter ${\thetav_{s} \coloneqq \big( \theta_{s}(0),	\theta_{s}(1), \dots, \theta_{s}(m-1) \big)}$, where $\theta_{{s}}(j) \coloneqq P(j|{s})$. The \textit{parameter vector} 	$\Theta \coloneqq \Theta_{\mathcal{S}} \coloneqq \{ \thetav_{s} : {s}\in\mathcal{S} \}$ groups all parameters in context set $\mathcal{S}$. Therefore, the Markov chain is completely characterised by the couple $(\mathcal{S},\Theta_{\mathcal{S}})$. We use $\mathcal{C}_D$ to denote the class of all context sets of order up to $D$, and we define $L_D(\mathcal{S}) \coloneqq |\{s\in\mathcal{S}:\ l(s) = D\}|$ the number of contexts with length $D$. 

\begin{figure}[!t]
	\centering
	\includegraphics[width=0.4\linewidth]{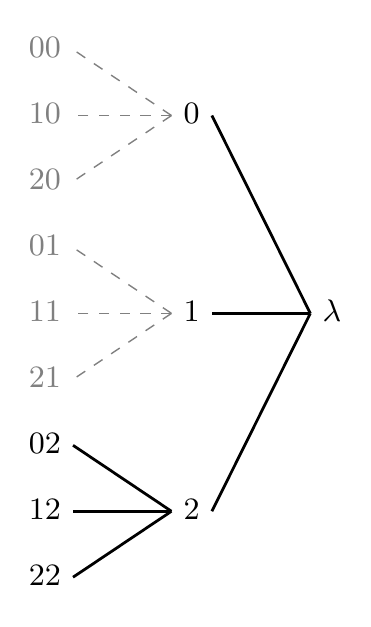}
	\caption{Example of model $\mathcal{S} = \{ 0, 1, 02, 12, 22 \} \subseteq \mathcal{T}_D$ with $m=3$ and $D=2$. In this case, we have $L_D(\mathcal{S}) = |\{ 02, 12, 22 \}| = 3$. The suffix of the sequence $x_{-\infty}^0 = \cdots 01$ is $c(x_{-\infty}^0) = c(01) = 1$. The (conditional) probability of the string $x_1^5 = 02212$ given the past symbols $x_{-1}^0 = 01$ is $P(x_1^5|x_{-1}^0) = \theta_1(0) \cdot \theta_0(2) \cdot \theta_{02}(2) \cdot \theta_{22}(1) \cdot \theta_{1}(2)$. After processing this sequence, the counter for context $s=1$ is $\pmb{a}_1(x_1^5) = \left(a_1(0), a_1(1), a_1(2)) = (1,0,1\right)$.}
	\label{fig:context-tree-example}
\end{figure}

Since the context set $\mathcal{S}$ is proper, its elements can be represented as leaf nodes of a tree $\mathcal{T}_D \supseteq \mathcal{S}$, called \textit{context-tree}. For a given sequence $x_1^n$, each leaf node $s \in \mathcal{S}$ is associated with a \textit{counter} $\pmb{a}_{s}~\coloneqq~\pmb{a}_{s}(x_1^n)~\coloneqq~{\big({a}_{s}(0), {a}_{s}(1), \dots, {a}_{s}(m-1)\big)}$, where ${a}_{{s}}(j)$ stores the number of times that symbol $j \in \mathcal{A}$ follows context ${s}$ in ${x}_{1}^{n}$. The counter of each inner node of the tree is recursively defined as the sum of the counters of its children nodes, i.e., $\av_s \coloneqq \sum_{j\in\mathcal{A}} \av_{js}$, $\forall\,s\in\mathcal{T}_D \setminus \mathcal{S}$. We use the empty string $\lambda$ to denote the root of the tree. An illustrative example of these concepts is given in Fig.~\ref{fig:context-tree-example}.

With the above definitions and the Markov property for a $D$-th order Markov chain, if both $\mathcal{S}$ and $\Theta_{\mathcal{S}}$ are known, the probability of a sequence can be written, as in~\cite{kontoyiannis},
\begin{align*} \label{eq:markov-property}
	P({x}_{1}^{n}|{x}_{D-1}^{0},\mathcal{S},\Theta_{\mathcal{S}}) &= \prod_{i=1}^{n}P(x_i|{x}_{i-D}^{i-1},\mathcal{S},\Theta_{\mathcal{S}}) \nonumber\\	
	&= \prod_{i=1}^{n}P(x_i|c({x}_{i-D}^{i-1}),\mathcal{S},\Theta_{\mathcal{S}}) \nonumber\\
	&= \prod_{i=1}^{n}\theta_{c({x}_{i-D}^{i-1})}(x_i) 
	= \prod_{{s}\in\mathcal{S}} \prod_{j\in\mathcal{A}} {\theta_{s}(j)}^{{a}_{s}(j)},
\end{align*}
as a simple function of the parameters $\mathcal{S}$, $\Theta_{\mathcal{S}}$, as well as the counters of the leaf nodes.

If only the model $\mathcal{S}$ is known, but not its parameters $\Theta_{\mathcal{S}}$, the \textit{marginal distribution} of a sequence ${x}_{1}^{n}$, given its past ${x}_{1-D}^{0}$ and model $\mathcal{S}$, is
\begin{equation}
P({x}_{1}^{n}|{x}_{1-D}^{0},\mathcal{S}) =
\int P({x}_{1}^{n}|{x}_{1-D}^{0},\mathcal{S},\Theta)\pi(\Theta|\mathcal{S})\ \mathrm{d}\Theta,
\label{eq:S_known}
\end{equation}
assuming the distribution of the parameters $\pi(\Theta|\mathcal{S})$ is known. While this distribution is unknown in general, using the so-called \textit{Jeffrey's prior} is asymptotically optimal in the minimax sense~\cite{gassiat}. This choice corresponds to setting $\pi(\Theta|\mathcal{S})$ to be the Dirichlet distribution with parameters $\left(\frac{1}{2}, \cdots, \frac{1}{2}\right)$. It turns out that, under this assumption, the marginal distribution \eqref{eq:S_known} can be simplified to the so-called \emph{Krichevsky–Trofimov~(KT) distribution}, which can be easily computed as	
\begin{equation} \label{eq:kt-estimator}
	P({x}_{1}^{n}|{x}_{1-D}^{0},\mathcal{S}) 
	= \prod_{{s} \in \mathcal{S}} P_e(\pmb{a}_{s}),
\end{equation}
where
\begin{equation} \label{eq:Pe}
	P_{e}(\pmb{a}_{s}) =
	\frac{\prod_{j \in \mathcal{A}}
		\left(\frac{1}{2}\right)\left(\frac{3}{2}\right)\cdots\left(\pmb{a}_{s}(j)-\frac{1}{2}\right)}{\left(\frac{m}{2}\right)\left(\frac{m}{2}+1\right)\cdots\left(\frac{m}{2}+M_{s}-1\right)},
	\quad s\in\mathcal{T}_D, 
\end{equation}
with $M_{s} \coloneqq \sum_{j=0}^{m-1}\pmb{a}_{s}(j)$. We note that other prior distributions $\pi(\Theta|\mathcal{S})$ have been considered in the literature and lead to different marginal distributions~\cite{tjalkens-97,volf}.

Finally, if the model $\mathcal{S}$ is also unknown, then we shall marginalise over $\mathcal{S}$ with a given prior distribution $\pi_D$ on all models $\mathcal{S}$ of maximal depth $D$. Fixing $\gamma\in\left]0,1\right[$ and
\begin{equation} \label{eq: prior}
	\pi_D(\mathcal{S}) \coloneqq
	{(1-\gamma)}^{\frac{|\mathcal{S}|-1}{m-1}}{\gamma}^{|\mathcal{S}|-L_D(\mathcal{S})},
\end{equation}
we obtain a mixture of different distributions \eqref{eq:kt-estimator}, corresponding to the coding distribution of CTW~\cite{gassiat, kontoyiannis}:
\begin{equation}
	Q_n(x_1^n | x_{1-D}^0) \coloneqq \sum_{\mathcal{S}\in \mathcal{C}_D}
	\pi_D(\mathcal{S})  \prod_{{s} \in \mathcal{S}} P_e(\pmb{a}_{s}). 
\end{equation}

Not only is this coding distribution universal for the class of stationary ergodic sources, but also it can be recursively computed so that complexity is linear in~$n$. The essence of the context-tree weighting~(CTW) algorithm is based on the following definitions and results.

\begin{figure*}[!b]
	\normalsize
	\setcounter{tempeqcounter}{\value{equation}}
	\setcounter{equation}{11}
	\vspace*{4pt}
	\hrulefill
	\begin{equation} \label{eq:maximising-tree}
	\mathcal{S}_m^{s} \coloneqq \left\{\begin{array}{ll}
	\bigcup_{j=0}^{m-1}\mathcal{S}_m^{j{s}}\times \{j\}, &\quad \text{if}\ (1-\gamma)\prod_{j=0}^{m-1}P_m^{j{s}} > P_e(\pmb{a}_{s})\ \text{and}\ 0 \le d < D,\\
	\{\lambda\}, &\quad \text{if}\ (1-\gamma)\prod_{j=0}^{m-1}P_m^{j{s}} \le P_e(\pmb{a}_{s})\ \text{and}\ 0 \le d < D,\\
	\{\lambda\}, &\quad \text{if}\ d=D.
	\end{array}\right.
	\end{equation}
	\setcounter{equation}{\value{tempeqcounter}+1}
\end{figure*}

\begin{definition} \label{def:weighting-probability}	
	For $\gamma \in\left]0,1\right[$, to each node ${s}\in\mathcal{T}_D$, with $l(s)=d$, we assign a \emph{weighted probability} $P_{w}^{{s}}$, defined as
	\begin{equation}
	P_{w}^{{s}} \coloneqq
	\begin{cases}
	\gamma P_e(\pmb{a}_{{s}}) + (1-\gamma)\prod_{j=0}^{m-1}P_w^{j{s}}, & 0 \le d < D,\\
	P_e(\pmb{a}_{{s}}), & l({s}) = D.
	\end{cases}
	\end{equation}
	The context-tree together with the weighted probabilities of the nodes is called \emph{weighted context-tree}.
\end{definition}

\begin{lemma}[See~\cite{willems-ctw, kontoyiannis}] \label{lemma:weighting-general}	
	The weighted probability $P_w^{\lambda}$ of the root node ${\lambda}\in\mathcal{T}_D$ satisfies
	\begin{equation}
	P_w^{\lambda} = \sum_{\mathcal{S}\in\mathcal{C}_{D}} \pi_D(\mathcal{S}) \prod_{s\in\mathcal{S}} P_e(\pmb{a}_{s}).
	\end{equation}
\end{lemma}

This lemma shows that the CTW probability $Q_n(x_1^n | x_{1-D}^0)$ is indeed the weighted probability $P_w^\lambda$ of the root node $\lambda$. Therefore, to compute the CTW probability of $x_{1}^n$, the CTW algorithm updates $P_w^s$ sequentially on $x_1,\ldots,x_1^i,\ldots,x_1^n$. Details are omitted and can be found in~\cite{willems-ctw}.

A modification of the CTW algorithm yields the context-tree maximising~(CTM) algorithm~\cite{willems-ctm}, which computes the maximum \textit{a posteriori} model for a given sequence.

\begin{definition} \label{def:general-maximised-probability}	
	For $\gamma \in\;]0,1[$, to each node $s \in \mathcal{T}_D$, with $l(s)=d$, we assign a \emph{maximised probability} $P_m^{s}$, defined as
	\begin{equation} \label{eq:Pm}
	P_{m}^{{s}} \coloneqq
	\begin{cases}
	\max\{\gamma P_e(\pmb{a}_{{s}}),
	(1-\gamma)\prod_{j=0}^{m-1}P_m^{j{s}}\}, & 0 \le l({s}) < D\\
	P_e(\pmb{a}_{{s}}), & l({s}) = D.
	\end{cases}
	\end{equation}
	The \emph{maximising set} $\mathcal{S}_m^{s}$ is computed as shown in~\eqref{eq:maximising-tree}.
	The context-tree, together with the maximised probability distribution and the maximising sets, is called \emph{maximised context-tree}.
\end{definition}

\setcounter{equation}{12}

\begin{lemma}[See~\cite{willems-ctm, kontoyiannis}] \label{lemma:maximising-general}
	The maximised coding distribution $P_m^{s}$ of the root node
	${\lambda}\in\mathcal{T}_D$ satisfies
	\begin{equation}
	P_m^\lambda
	= \pi_{D}(\mathcal{S}^{\lambda}_m) \prod_{{s}\in\mathcal{S}_m^\lambda} P_e(\pmb{a}_{s})
	= \max_{\mathcal{S}\in\mathcal{C}_{D}} \pi_{D}(\mathcal{S}) \prod_{s\in\mathcal{S}} P_e(\pmb{a}_{{s}}).
	\end{equation}
\end{lemma}

It follows that the maximising set $\mathcal{S}_m^{\lambda}$, which is associated to the maximising probability $P_m^\lambda$, corresponds to the maximum \textit{a posteriori} model:
\begin{align*}
	\mathcal{S}_m^{\lambda}
	&= \arg\max_{\mathcal{S}\in\mathcal{C}_D}P(\mathcal{S}|{x}) 
	= \arg\max_{\mathcal{S}\in\mathcal{C}_D} \frac{\pi_D(\mathcal{S}) P({x}|\mathcal{S})}{P({x})} \nonumber\\
	&= \arg\max_{\mathcal{S}\in\mathcal{C}_D} \pi_D(\mathcal{S}) \prod_{{s}\in\mathcal{S}} P_e(\pmb{a}_{s}).
\end{align*}

Proof of Lemmas \ref{lemma:weighting-general} and \ref{lemma:maximising-general} were given, for the special case $m = 2, \gamma = 1/2$, in~\cite{willems-ctw} and~\cite{willems-ctm}, respectively, while the general case is addressed in~\cite{kontoyiannis}.

\section{Quantiser Design}\label{sec:quantisation}

\begin{figure}[!t]
	\centering
	\includegraphics[width=0.9\linewidth]{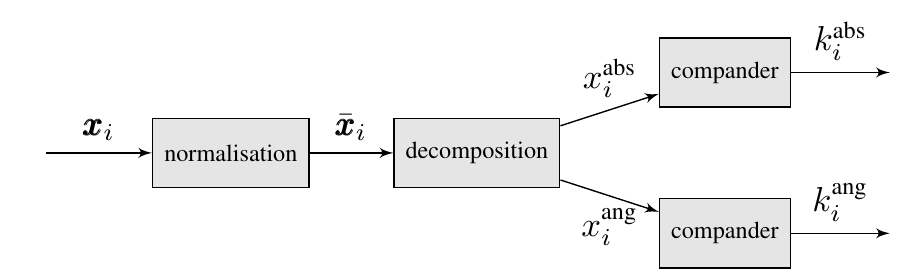}
	\caption{Block diagram for vector quantisation.}
	\label{fig:quantisation-diagram}
\end{figure}

The proposed vector quantisation consists in vector normalisation, decomposition into real components, and individual scalar quantisation based on parametric companders, as indicated in Fig.~\ref{fig:quantisation-diagram}. In the following, we elaborate each step.

\subsection{Vector Normalisation} \label{subsec:normalisation}

In this step, the input vector $\xv\coloneqq \left(x(1), \dots, x(\Nt)\right)$ is normalised by the component with the largest absolute value, i.e., $\bar{\xv} = \xv/x(i^*)$ where $i^* \coloneqq \arg\max_{i \in [\Nt]} |x(i)|$. Note that $\bar{x}(i^*) = 1$, while the other normalised components are complex in general with absolute value in $[0,1]$. The $i^*$-th component can skip the following steps and be directly assigned a special index indicating it as the strongest component. Because of this special symbol, our compression alphabet sizes $m$ have one more element than the number of quantisation levels $M$, i.e., $m = M + 1$.

\subsection{Decomposition}

Before scalar quantisation, each complex component has to be decomposed into real values. Two straightforward options are: 1)~Cartesian decomposition into real real and imaginary parts, and 2)~polar decomposition into amplitude and phase. We consider the polar decomposition, since the amplitude and phase are usually less correlated in wireless applications, therefore providing a less `redundant' representation of the bounded complex number. Indeed, the real and imaginary parts of the normalised complex components tend to be correlated, e.g., strong real part implies weak imaginary part since the amplitude is bounded by $1$.

\subsection{Quantisation with Parametric Companders}

The amplitude and phase are separately quantised with different scalar quantisers of $M_{\abss}$ and $M_{\ang}$ quantisation levels, respectively. 

If the input symbols of the quantiser are uniformly distributed, then a uniform quantiser is optimal. In general, however, uniform quantisation can be far from optimal in the rate-distortion sense~\cite{cover}. Let~$\rv{X}$ be a random variable representing the input, following some distribution $P$ with support~$[0,1]$. The idea of using a compander~\cite{bennett} is to apply a non-linear and non-decreasing mapping~$g: [0,1] \to [0,1]$ to the signal (\emph{compression}) before quantising it, so that the signal is more `uniform' in the image space. To recover the signal, the inverse mapping $g^{-1}: [0,1] \to [0,1]$ is used (\emph{expansion}); hence the name \emph{compander}. It is practical to use parametric companders, i.e., differentiable mappings $g$ that can be described by a few number of parameters, such as the $\mu$-law compander, characterised by a single parameter $\mu>0$. Note that, as compared to the Lloyd quantiser~\cite{cover}, compander-based quantisers have much lower complexity of quantisation and representation.

We propose a data-driven design for companders parametrised by some $\xi$~(which can contain multiple scalar parameters). Assume that we have a set of training data~$\{x_1,\ldots,x_n\}$. Our design follows a two-step procedure:~1)~uniformisation of the data, and 2)~adjustment of the compander parameters, as follows.

\subsubsection{Uniformisation of the Data}

We assume that the training data are independent samples from some distribution~$P$. If we knew the cumulative distribution function~(cdf) $F_P$ of $P$, we could apply the mapping $F_P$ such that $\{F_P(x_1),\ldots,F_P(x_n)\}$ are samples from a uniform distribution. If, however, we are restricted to a class of companders~$\{g_\xi \colon \xi\in\Xi\}$, for some set~$\Xi$, then we have to approximate  $F_P$ by $g_\xi$. Since a compander, as defined above, is non-decreasing from~$0$ to~$1$, it is equivalent to a cdf. Thus, a sensible criterion for the approximation is through the Kullback-Leibler divergence:
\begin{align} 
	\xi^* &= \arg\min_{\xi \in \Xi} D(P \, \| \, g_\xi) \nonumber\\
	&= \arg\min_{\xi \in \Xi} \left\{-H(\rv{X}) - \mathbb{E}_P[\log (g'_\xi(\rv{X}))] \right\} \nonumber\\
	&= \arg\max_{\xi \in \Xi} \mathbb{E}_P[\log (g'_\xi(\rv{X}))].
	\label{eq:max-qn}
\end{align}
Interestingly, this is equivalent to maximising the differential entropy of $g_\xi(\rv{X})$. As the uniform distribution maximises differential entropy among all bounded support distributions~\cite{cover}, the criterion~\eqref{eq:max-qn} indeed returns the best `uniformiser'. Note that since $g_\xi$ is a cdf, $g'_\xi$ is the corresponding probability density function~(pdf).

The true distribution of the data is, however, unknown in most practical scenarios. But we can adapt the probabilistic criterion~\eqref{eq:max-qn} into a data-driven one by replacing the expectation with the sample mean: 
\begin{equation} \label{eq:max-qn2}
\arg\max_{\xi \in \Xi} \frac{1}{n} \sum_{i=1}^n \log (g'_\xi(x_i)).
\end{equation}

In this paper, we consider the $\mu$-law compander and another one that we call $\beta$-law compander, as shown in Table~\ref{tab:distributions}. The $\beta$-law compander is equivalent to the beta cdf, parametrised by $\alpha>0$ and $\beta>0$. An attractive feature of the $\beta$-law compander is that the corresponding pdf is log-concave in $(\alpha,\beta)$~\cite[Theorem~6]{dragomir}, so that the maximisation \eqref{eq:max-qn2} is concave and can thus be easily solved. 

\begin{table*}[!t]
	\centering
	\caption{Some Compander Functions.}
	\label{tab:distributions}
	\renewcommand{\arraystretch}{1.2}
	\tabcolsep=12pt
	\begin{tabular}{cccc}
		\hline
		Compander type & Parameters $\xi$ & cdf  $g_{\xi} (x)$ & pdf $g_{\xi}'(x)$ \\ 
		\hline \hline
		\rule{0pt}{5ex}
		$\mu$-law & $\mu > 0$ & $\dfrac{\ln(1 + \mu x)}{\ln(1 +\mu)}$ & $\dfrac{\mu}{(1+\mu x) \ln(1 + \mu)}$ \\
		\rule{0pt}{5ex}
		$\beta$-law  & $\alpha>0,\ \beta>0$ & $\dfrac{\Gamma(\alpha+\beta)}{\Gamma(\alpha)\Gamma(\beta)}\displaystyle\int_{0}^{x} t^{\alpha-1}(1-t)^{\beta-1}\ \mathrm{d}t $ & $\dfrac{\Gamma(\alpha+\beta)}{\Gamma(\alpha)\Gamma(\beta)} x^{\alpha-1}(1-x)^{\beta-1}$ \vspace{1mm} \\
		\hline
	\end{tabular}
	\vspace*{8pt}
\end{table*}

\subsubsection{Adjustment of the Compander Parameters}

\begin{figure}[!t]
	\centering
	\includegraphics[width=0.8\linewidth]{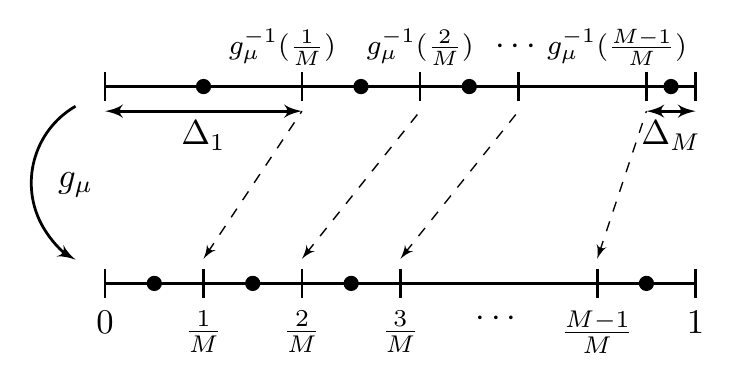}
	\caption{$\mu$-law compander quantiser of size $M$.}
	\label{fig:mu-law}
\end{figure}

Note that uniformising the input is not enough in the sense of rate-distortion. If we perform uniform quantisation right after this step, then the uniformisation only makes the number of samples in each quantisation interval as uniform as possible. If the number of samples are exactly the same in each interval, the average distortion is dominated by the one generated in the largest interval---we illustrate this argument in Fig.~\ref{fig:mu-law} with a $\mu$-law compander. Therefore, we need to adjust the parameter to balance the distortion generated in different intervals, which is the role of the second step. While the exact solution is hard to find, we provide a heuristic, yet efficient way to make the adjustment.

\begin{figure}[!t]
	\centering
	\subfloat[Amplitude.]{\includegraphics[width=0.5\linewidth]{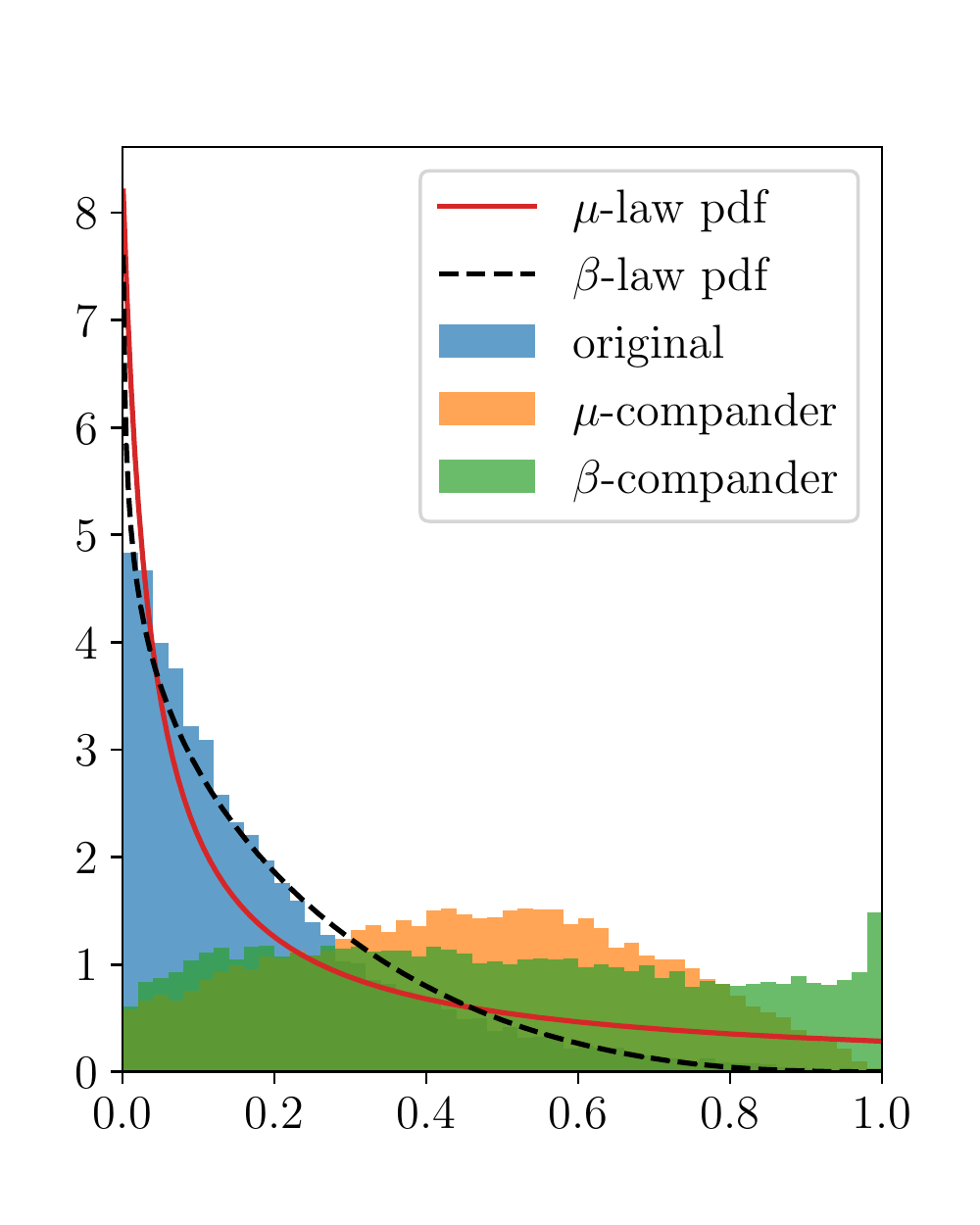}}
	\hfil
	\subfloat[Phase.]{\includegraphics[width=0.5\linewidth]{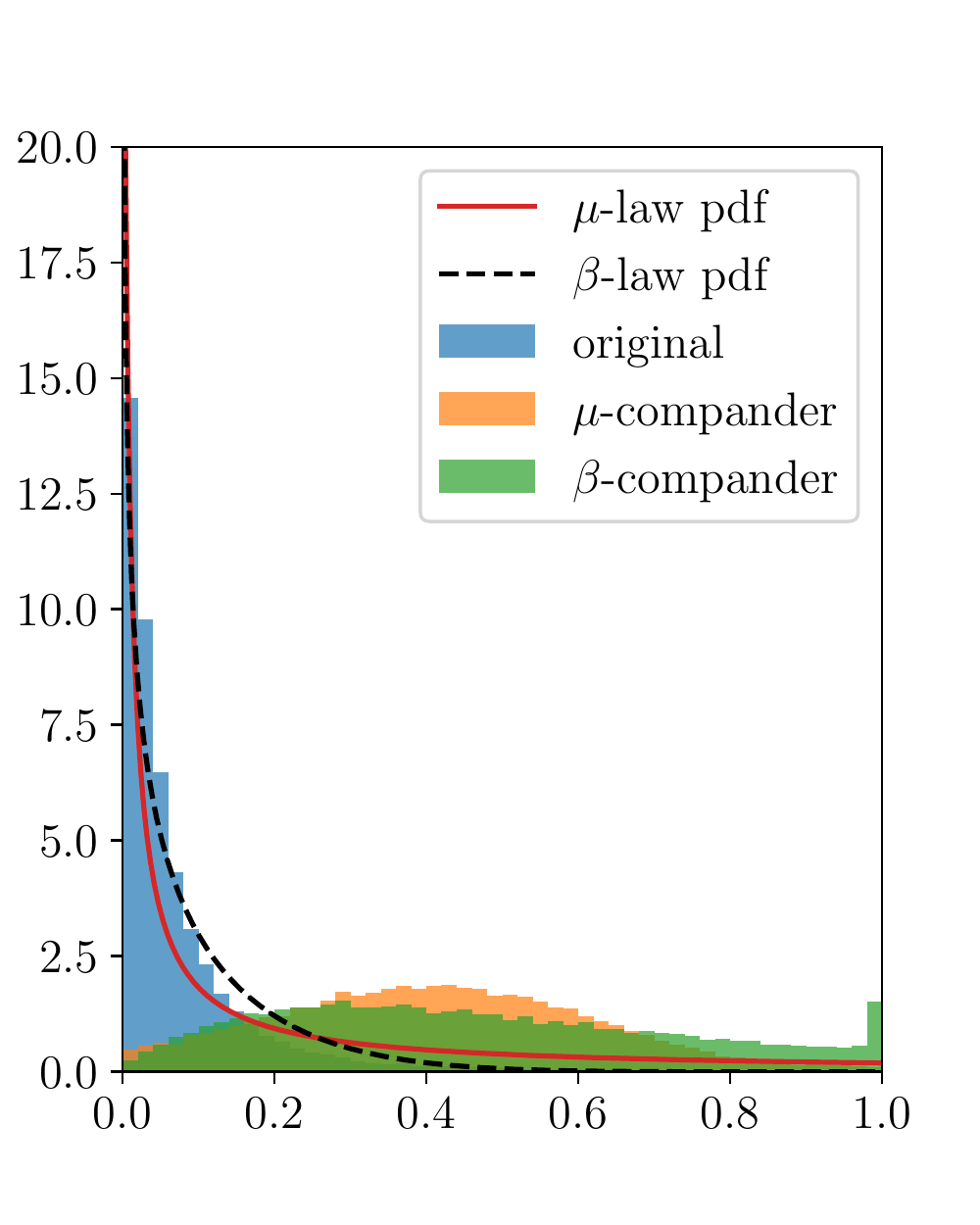}}
	\caption{Normalised histograms from CSI vector sequences (EVA70, high correlation, cf.~Section~\ref{sec:results}) before and after applying $\mu$-law and $\beta$-law companders.}
	\label{fig:histograms}
\end{figure}

Consider a quantiser with $M$ levels. If we assume that the distortion generated in the $i\text{-th}$ interval is proportional to the squared length $\Delta_i^2$ of that interval, then the average distortion is proportional to $\sum_{i=0}^{M-1} N_i \Delta_i^2$, where $N_i$ is the number of samples inside the $i\text{-th}$ interval. Here, each interval~$i$ contributes with $N_i \Delta_i^2$. Starting with the solution given by step~1, all $N_i$'s are similar, and the largest interval contributes the most in the average distortion; similarly, the smallest interval contributes the least. The idea is therefore to reduce the largest interval until $ N_S \Delta_S^2 \ge N_L \Delta_L^2 $, where `S' and `L' stand for the `smallest' and `largest' intervals, respectively. For instance, with the $\mu$-law compander, the smallest interval is always the first one and the largest, the last one. Reducing $\mu$ towards $0$ would reduce the gap between the extreme interval sizes. With the $\beta$-law compander, the smallest and largest intervals depend on the parameters $(\alpha,\beta)$, but letting $(\alpha,\beta)$ go towards $(1,1)$ also reduces the gap between the extreme interval sizes. In Fig.~\ref{fig:histograms}, we plot the histogram of some data from CSI vectors and the output of the two companders. We see that the histogram is indeed flattened after applying the companders.

Although the presented compander design is based on training data, we could also start with a uniform compander and update it regularly when more data is available. A great advantage of the parametric compander design is the negligible communication overhead of the (few) quantisation parameters.

\subsection{Quantisation Resolutions for Amplitude and Phase}

Since we quantise the amplitude and phase of a complex symbol separately, we would like to find out the `optimal' resolutions for both quantisers. Let $M_{\abss}$ and $M_{\ang}$ be the respective number of quantisation levels, then $M = M_{\abss} M_{\ang}$ is the total number of quantisation levels for a complex number. The question is thus to find out the optimal values, $M_{\abss}^*$ and $M_{\ang}^*$, for a given $M$. While the exact solution would depend on the distribution of the complex number, we are interested here in finding a rule of thumb based on sensible assumptions. We can show that the optimal solution is such that the respective quantisation errors satisfy $\epsilon_{\abss} \approx \mathbb{E}\left[ |\rv{X}|^2 \right] \epsilon_{\ang}$, where $\rv{X}$ is the complex input. 

The problem is formulated as an optimisation of the MSE of the complex variable $\rv{X} = A e^{\mathbf{j}\Phi}$, i.e., to minimise $\epsilon \coloneqq \mathbb{E}\bigl[ |A e^{\mathbf{j}\Phi} - \hat{A} e^{\mathbf{j}\hat{\Phi}}|^2 \bigr]$, where $A \in \left[0,\infty\right[$ and $\Phi\in\left[0,2\pi\right[$. Letting $\tilde{A} \coloneqq A -\hat{A}$ and $\tilde{\Phi} \coloneqq \Phi - \hat{\Phi}$, we have $\epsilon = \mathbb{E}\bigl[ \tilde{A}^2 \bigr] + 2\mathbb{E}\bigl[ A \hat{A} (1-\cos(\hat{\Phi})) \bigr]$. Let us reasonably assume that 1)~$\mathbb{E}\bigl[ \tilde{A} \bigr] = 0$; 2)~$\hat{A}$ and $\tilde{A}$ are uncorrelated; and 3)~$A\hat{A}$ and $1-\cos(\tilde{\Phi})$ are uncorrelated. Then, using the approximation~${1-\cos(x) \approx {x^2}/{2}}$, we have
\begin{align}
	\epsilon &\approx \mathbb{E}\bigl[ \tilde{A}^2 \bigr] + \mathbb{E}\bigl[ \hat{A}^2 \bigr] \mathbb{E}\bigl[ \tilde{\Phi}^2 \bigr] = \epsilon_{\abss} + (E_{\abss} - \epsilon_{\abss}) \epsilon_{\ang}, \label{eq:epsilon_app}
\end{align}%
where ${E_{\abss} \coloneqq \mathbb{E}\bigl[ {A}^2 \bigr] =\mathbb{E}\bigl[ |\rv{X}|^2 \bigr]}$, ${\epsilon_{\abss} \coloneqq \mathbb{E}\bigl[ \tilde{A}^2 \bigr]}$, and ${\epsilon_{\ang} \coloneqq \mathbb{E}\bigl[ \tilde{\Phi}^2 \bigr]}$. Since the regime of interest is such that $\epsilon_{\ang} \le 1$, the approximation~\eqref{eq:epsilon_app} is an increasing function of $\epsilon_{\abss}$ and of $\epsilon_{\ang}$. Intuitively, the overall quantisation error is increasing with the individual quantisation errors.

To finally find the optimal $M_{\abss}^*$ and $M_{\ang}^*$ that minimise \eqref{eq:epsilon_app}, we make the following `mild' assumption: both $\epsilon_{\abss}$ and $\epsilon_{\ang}$ decrease with $M_{\abss}$ and $M_{\ang}$ as
\begin{equation}
	\epsilon_{\abss}/E_{\abss} \approx c_{\abss} M_{\abss}^{-2}, \quad 
	\epsilon_{\ang}/E_{\ang} \approx c_{\ang} M_{\ang}^{-2}, 
	\label{eq:RD}
\end{equation}
where $E_{\ang} \coloneqq \mathbb{E}\left[ \Phi^2 \right]$, and $c_{\abss}, c_{\ang}$ are constants depending on the respective marginal law of the normalised amplitude and phase.  This assumption is supported by the rate-distortion theorem~\cite{cover} and is in general observed for medium to high rate quantisers. Plugging in the constraint $M_{\abss} M_{\ang} = M$, relaxed from integers to all positive numbers, we obtain the constraint on $\epsilon_{\abss}$ and $\epsilon_{\ang}$
\begin{equation}
	\epsilon_{\abss} \epsilon_{\ang} \approx c_{\abss} c_{\ang} M^{-2}, 
	\label{eq:constraint_e}
\end{equation}
i.e., the product of $\epsilon_{\abss}$ and $\epsilon_{\ang}$ is a constant. Under this constraint, minimising \eqref{eq:epsilon_app} is equivalent to minimising $\epsilon_{\abss} + E_{\abss} \epsilon_{\ang}$ subject to $\epsilon_{\abss} \epsilon_{\ang} = \text{const}$. It can be readily shown that the optimal solution is such that $\epsilon_{\abss}^* = E_{\abss} \epsilon_{\ang}^*$ for any constant. From \eqref{eq:RD}, we see that 
\begin{equation}
	M_{\ang}^* \approx M_{\abss}^* \sqrt{E_{\ang} c_{\ang} /c_{\abss}}. 
\end{equation}

For example, if the normalised amplitude $A/\sqrt{E_{\abss}}$ and phase $\Phi/\sqrt{E_{\ang}}$ have the same distribution~(with bounded support), i.e., $c_{\abss} = c_{\ang}$, then $M_{\ang}^* \approx M_{\abss}^* \sqrt{E_{\ang}}$. If, in addition, they are both uniformly distributed, then $E_{\ang} = \frac{4}{3}\pi^2$, and $\sqrt{E_{\ang}} \approx 3.6$. Therefore, a rule of thumb for the number of quantisation bits is to use two more bits for the phase than for the amplitude.

\begin{remark}
	It is well known that, followed by entropic encoding, a uniform quantiser is asymptotically optimal in the high-rate regime. We emphasise, however, that we do not operate here in the high-rate regime, unlike many other applications. More importantly, a large alphabet size would make the following context-tree-based compression highly inefficient. Hence, a carefully designed quantiser is crucial for the overall performance.
\end{remark}

\section{Compression Algorithm}\label{sec:compressor}

\begin{figure}[!t]
	\centering
	\includegraphics[width=0.9\linewidth]{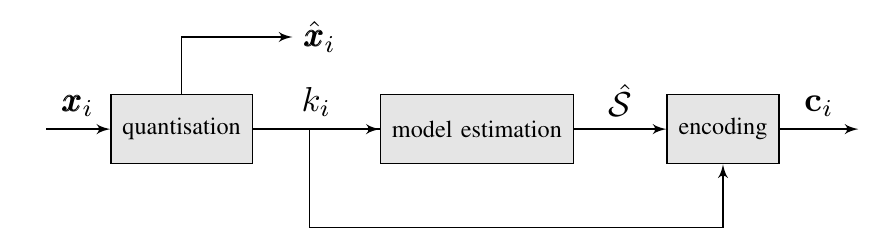}
	\caption{Block diagram for vector processing.}
	\label{fig:general-diagram}
\end{figure}

In order to compress the sequence of quantisation indices, one option is to directly apply CTW algorithm with arithmetic coding. In this section, we describe an alternative context-tree-based solution that limits the fluctuation of the output bit-stream. It consists in first estimating a tree model $\hat{\mathcal{S}}$, and then using the probabilities derived from that model to encode each symbol, as depicted in Fig.~\ref{fig:general-diagram}.

\subsection{Tree Model Estimation}

Given a scalar sequence $k_1^n$ of quantisation indices from an alphabet $\mathcal{A} = \{ 0, 1, \dots, m-1 \}$, we use the CTM algorithm~(cf. Section~\ref{subsection: markov-chain}) to find the maximum \textit{a posteriori} tree model $\hat{\mathcal{S}}$ that describes that sequence. This algorithm consists in building the same tree $\mathcal{T}_D$ as in CTW algorithm, followed by a pruning procedure, as described by \eqref{eq:maximising-tree}. Both the computational and storage complexity of CTM algorithm are known to be $O(nmD)$, i.e., linear with sequence length $n$, alphabet size $m$ and maximum tree depth $D$, cf.~\cite{kontoyiannis}.

When training data are available, we can apply the CTM algorithm on the training data to estimate the MAP model $\hat{\mathcal{S}}$, and use it to estimate symbol probabilities and encode the incoming sequence. This, however, is not mandatory: we can also initialise the full tree $\mathcal{T}_D$ with empty counts, keep updating them with incoming data, and regularly prune a copy of this tree to have an updated estimate of the MAP model~$\hat{\mathcal{S}}$. Similarly, if the sequence is not stationary, we can also make the model forget symbols it has seen in a distant past by decreasing the counts of the tree model $\mathcal{T}_D$. With this, at each instant, the model is built upon the observation of a sequence of only the most recent symbols. This can be done without increasing the complexity of the algorithm.

\subsection{Prediction and Encoding}

Once a tree model $\hat{\mathcal{S}}$ is estimated, we can encode a sequence $k_1^n$ according to the probabilities issued from that model. Note that, given a model $\hat{\mathcal{S}}$ and past symbols $k^{0}_{1-D}$, the estimated probability of $k_1^n$ can be computed via the KT estimator, using \eqref{eq:kt-estimator} and \eqref{eq:Pe}. In particular, denoting $s \coloneqq c(k_{i-D}^{i-1})$, we can compute the probabilities $\hat{P}(\cdot) \coloneqq P(\cdot | \hat{\mathcal{S}})$ that the next symbol is $k_i=j$, for all $j \in \mathcal{A}$, as
\begin{align} \label{eq:P-next}
	\hat{P}(j | {k}_{i-D}^{i-1}) 
	&= \frac{\hat{P}({k}_{i-D}^{i})}{\hat{P}({k}_{i-D}^{i-1})}
	=\frac{\prod_{{s}'\in\hat{\mathcal{S}}}P_e(\pmb{a}_{{s}'}({k}_1^{i}))}{\prod_{{s}'\in\hat{\mathcal{S}}}P_e(\pmb{a}_{{s}'}({k}_{1}^{i-1}))} \nonumber\\
	&= \frac{P_e(\pmb{a}_{{s}}({k}_{1}^{i}))}{P_e(\pmb{a}_{{s}}({k}_{1}^{i-1}))}
	= \frac{a_{{s}}(j)+\frac{1}{2}}{\frac{m}{2}+\sum_{j' \in \mathcal{A}} a_{{s}}(j')}.
\end{align}
With $\hat{P}$, one could apply arithmetic coding to encode $k_i$. But the encoded bit-stream would have a variable length depending on both $\hat{P}$ and $k_i$, and reducing the fluctuation of the coded bit length is important in practical communication systems. On the other extreme, a fixed length coding does not exploit the knowledge of the coding distribution $\hat{P}$ and does not compress at all. Here, we propose an encoding scheme with three possible codeword lengths.

We assume that both encoder and decoder keep a synchronised version of the tree. In addition, we use an auxiliary lower resolution quantiser to apply on least probable symbols. Fix two integers $q_1,q_2 \le \log m$ such that $m_1 \coloneqq 2^{q_1}$, $m_2 \coloneqq 2^{q_2}$, and $m_1 + m_2 \le m$. Each incoming symbol $k_i \in \mathcal{A}$ at instant $i$ is encoded as follows.
\begin{itemize}
	\item If $k_i$ is among the $m_1$ most probable symbols~(tie could be broken with a fixed rule) according to $\hat{P}$, then the encoded bit string is $\mathbf{c}_i = 0$ followed by $q_1$ bits indicating the position of $k_i$ in the list of the $m_1$ most probable symbols.
	
	\item Otherwise, if $k_i$ is among the next $m_2$ most probable symbols, the encoded bit string $\mathbf{c}_i$ is $10$ followed by $q_2$ bits indicating the position of $k_i$ in the second list.
	
	\item Finally, if $k_i$ is not among the $m_1+m_2$ most probable symbols, the encoded bit string is $\mathbf{c}_i$ is $11$ followed by $\lceil \log m_3 \rceil$~bits corresponding to the index $\tilde{k}_i$ from a lower resolution quantiser of $m_3-1$ levels.
\end{itemize}

Note that, with this scheme, the codeword length is either $1+q_1$, $2+q_2$, or $2+ \lceil \log m_3 \rceil$. The proposed scheme can be extended to more levels if desired. In the following, we fix $q_1=0$ so that $m_1=1$.

It is worth pointing out that the decoder does not have access to the original high-resolution index $k_i$ when the third case happens at time $i$. This prevents the decoder from normally updating its tree. But encoder and decoder must update the tree in the same way so that the codebooks remain synchronised at both sides. A workaround is to let both the encoder and the decoder update the tree with a projection of the low-resolution index $\tilde{k}_i$ back to the high-resolution codebook. Specifically, we reconstruct the vector using the low-resolution quantiser, re-quantise it with the high-resolution quantiser, and use the corresponding index. Another way could be to simply ignore this symbol for tree update.

We have two main reasons to consider this strategy. First, encoding a symbol and decoding a binary string can be immediately done. Furthermore, the length of the encoded bit-stream is within a fixed number of levels. This aspect is a difference from arithmetic encoding, in which the output is of variable length, which may lead to difficulties when implementing CSI feedback in real systems. In exchange, we cannot expect the asymptotically optimal two-bits redundancy enjoyed by that method.

\subsection{Multiple Trees} \label{subsec:multiple-trees}

In practice, we may want to compress many processes simultaneously, as in the application to CSI representation. Multiple trees come both from the decomposition of complex components into amplitude and phase, and from the fact that the BS has multiple antennas. While each tree provides the marginal distribution of the given sequence, all the marginal distributions can be jointly used to encode the parallel streams together, in order to improve the coding rate. The intuitive idea is that if the indices of all the processes at a given time instant agree with the prediction of the respective models, they need not be individually encoded. If, however, this is not the case, only information about the indices that differ from the model prediction need be transmitted.

Consider that, at each time instant, we have $\Nt$ (complex) processes, thus $2\Nt$ scalar symbols to compress (amplitude and phase indices). For the sake of explanation, let us introduce an auxiliary variable~(\emph{flag}) $\Delta_l$, defined as follows. For each symbol $k_l$ to be compressed, where, here, the index $l \in [2\Nt]$ denotes the process (and not time):
\begin{itemize}
	\item if $k_l$ is the most probable symbol~(tie could be broken with a fixed rule) according to $\hat{P}$, then $\Delta_l=0$;
	\item otherwise, if $k_l$ is among the next $2^{q_{l}}$ most probable symbols, $\Delta_l=1$;
	\item finally, if $k_l$ is not among the $1+2^{q_{l}}$ most probable symbols, then it is encoded with a lower resolution codebook of size $m_{L,l}$, and $\Delta_l=2$.
\end{itemize}

Note that, with this notation, the individual compression rate, described in the previous subsection, is given by~\eqref{eq:rate-indiv}.
\begin{align} \label{eq:rate-indiv}
	R_{\text{individual}} = \sum_{l=1}^{2\Nt} \Big( &\1_{\{\Delta_l=0\}} + \1_{\{\Delta_l=1\}}(2+q_l) \nonumber\\
	&+ \1_{\{\Delta_l=2\}}(2+\lceil \log m_{L,l} \rceil) \Big).
\end{align}

Now, the joint description is composed of two parts: the \emph{state indicator} that contains information about which of the $\Nt$ processes have `varied', i.e., did not follow the correspondent model prediction, and the \emph{change}, which represents the symbols that varied. Therefore the joint rate is written as the sum of the rates of the state indicator and the change parts:
\[ R_{\text{joint}} = R_{\text{indicator}} + R_{\text{change}}. \]

Two joint strategies are described in the following.

\subsubsection{Simple Strategy}

A simple way to encode the state indicator is as follows: if all sequences follow the model, encode that with a $0$. Otherwise, use $\Nt$ bits to indicate which processes (antennas) had some sequence (either amplitude or phase) that varied with respect to the model prediction. This requires
\begin{equation}
R_{\text{indicator}} = \left(1 + \Nt\1_{\{ \Delta_1 + \Delta_2 + \cdots + \Delta_{2\Nt} > 0 \}}\right) / \Nt
\end{equation}
bits per process.

To describe the variation with respect to the model prediction, we need
\[ R_{\text{change}} = C_{\abss} + C_{\ang}, \]
with $C_{\abss}$ and $C_{\ang}$ given by
\begin{align} \label{eq:E-abs-ang}
	C_{i} = \sum_{l=1}^{\Nt} \Big( &\1_{\{\tilde\Delta_l = 0\}}(1 + q_{i}) + \1_{\{\tilde\Delta_l = 1\}}(1 + q_{i}) \nonumber\\
	&+ \1_{\{\tilde\Delta_l=2\}}(1+\lceil \log m_{L,i} \rceil) \Big),
\end{align}%
where $i$ is either `abs' or `ang', each sum is over the $\Nt$ processes of the respective type (amplitude or phase), and $\tilde{\Delta}_l$ is the $\Delta_l$ of the corresponding type (amplitude or phase). In addition, $q_{i}$ is the number of bits needed to describe the list of most probable symbols of an amplitude or phase component, and $m_{L,i}$ is the lower resolution alphabet sizes for amplitude or phase.

Note that we have to encode both the variation of amplitude and phase, as soon as at least one of them varied, for a given antenna, since the indicator part only informs that a sequence has varied, but does not indicate which of them (amplitude or phase). Moreover, in addition to the number of bits needed to describe the symbol---either as the index in the list of most probable symbols or in the lower resolution alphabet---, an additional bit is needed to inform which of these cases has happened.

\subsubsection{Context-Tree Compression of State Indicator}

Alternatively, a more sophisticated strategy is to consider the sequence of state indicators. We consider the state indicator, an array of $\Nt$ bits, as an integer number between $0$ and $2^{\Nt} - 1$. Then, the sequence of state indicators can be itself compressed using the proposed context-tree-based method~(without the lower resolution codebook). In this case, the indicator is described with
\begin{equation}
	R_{\text{indicator}} = R_{\text{encoded}} / \Nt
\end{equation}
bits per process, where $R_{\text{encoded}}$ is the length of the binary output generated by the compression scheme.

The change part is as before, i.e.,
\begin{equation}
	R_{\text{change}} = C_{\abss} + C_{\ang},
\end{equation}
with $C_{\abss}$ and $C_{\ang}$ given by \eqref{eq:E-abs-ang}.

\section{Simulation Results} \label{sec:results}

\subsection{Simulation Setup}

We use the MATLAB LTE Toolbox~\cite{matlab-lte} to simulate LTE MIMO downlink channels. The model is configured according to the parameters in Table~\ref{tab:matlab-parameters}. In particular, we consider low~(EPA5, Doppler $5$~Hz), moderate~(EVA30, Doppler $30$~Hz) and high mobility~(EVA70, Doppler $70$~Hz) scenarios, with low or high correlation between antennas at the base station.  Other relevant parameters used for quantisation and compression are presented in Table~\ref{tab:simulation-parameters}.

\begin{table}[!t]
	\centering
	\renewcommand{\arraystretch}{1.2}
	\caption{Simulation Parameters for MATLAB LTE Toolbox.}
	\label{tab:matlab-parameters}
	\begin{tabular}{ccc}
		\hline
		Field                     & Parameter          & Value       \\ \hline \hline
		\multirow{3}{*}{Cell-wide settings}
		& RC   	             & R.12             \\ 
		& DuplexMode         & FDD              \\
		& TotSubframes       & 10               \\
		\hline
		\multirow{10}{*}{\begin{tabular}[c]{@{}c@{}}Propagation channel\\ configurations\end{tabular}}
		& NRxAnts            & 1                                  \\ 
		& MIMOCorrelation    & High, Low		                  \\ 
		& NormalizeTxAnts    & On                                 \\ 
		& DelayProfile       & EPA, EVA                           \\ 
		& DopplerFreq        & 5, 30, 70                          \\ 
		& InitTime 			 & 0 to 9.99                          \\ 
		& NTerms             & 16                                 \\ 
		& ModelType          & GMEDS                              \\ 
		& NormalizePathGains & On                                 \\
		& InitPhase          & Random                             \\
		\hline
		\multirow{2}{*}{\begin{tabular}[c]{@{}c@{}}Timing and\\ frequency offset\end{tabular}}
		& toffset       & 7                                   \\
		& foffset       & 0                                   \\
		\hline
	\end{tabular}
\end{table}

\begin{table}[!t]
	\centering
	\renewcommand{\arraystretch}{1.2}
	\caption{Simulation Parameters for Quantisation and Compression.}
	\label{tab:simulation-parameters}
	\begin{tabular}{ccc}
		\hline
		Parameter                          & Symbol         & Value  \\
		\hline \hline
		Total sequence length                         & $n$            & $10^4$ \\
		Channel signal-to-noise ratio                 & $\mathrm{SNR}$ & 30 dB  \\
		Tree maximum depth                            & $D$            & 2      \\
		Context-tree weighting coefficient            & $\gamma$       & $0.5$  \\
		Training sequence size (\% of total length)   & --     	       & 20\%   \\
		Interval between tree updates (in symbols)    & --  	       & 100    \\
		\hline
	\end{tabular}
\end{table}

\subsection{Quantiser Design}

First, we are interested in evaluating the performance of the quantiser design. We consider three quantisation schemes: the $\mu$-law compander, the $\beta$-law compander, and the cube-split quantiser~\cite{decurninge}. Interestingly, the cube-split quantiser can be regarded as a complex compander adapted to the distribution of normalised complex Gaussian vectors.

\begin{figure}[!t]
	\centering
	\includegraphics[width=0.9\linewidth]{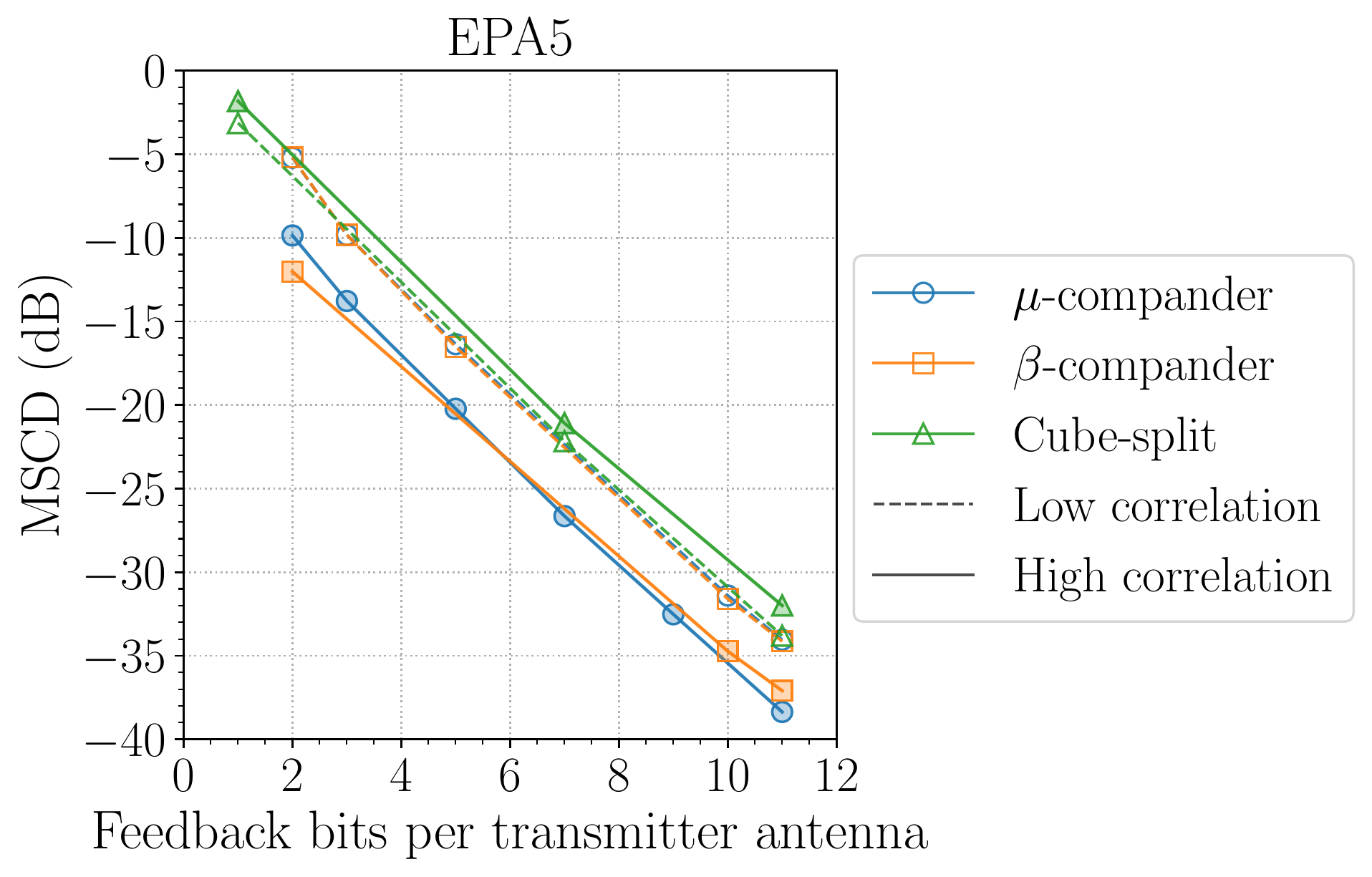}
	\caption{MSCD distortion for different quantisers.}
	\label{fig:quantiser-performance}
\end{figure}

In Fig.~\ref{fig:quantiser-performance} we plot the MSCD versus the feedback bit rate per antenna for the three quantisers, with no compression, for low and high antenna correlation, in the EPA5 scenario, with $\Nt=\Nr=4$. The plotted points correspond to the envelope formed by the best quantisation parameters (different codebook sizes) among those that were tested.

We see that, for low antenna correlation, the cube-split and the proposed quantisers achieve almost the same results. On the other hand, when antenna correlation is high, both proposed quantisers have similar performances and are noticeably better than the cube-split, which assumes uniformity of the distribution by design. The behaviour for EVA30 and EVA70 is similar and has been omitted.

\begin{remark}
	Although in this application we find that there is not much difference in using $\mu$-law or $\beta$-law compander, this may not be the case in other applications. The latter, having more adjustable parameters, could provide more flexibility in fitting experimental distributions.
\end{remark}

\subsection{Compression Algorithm}
\begin{figure*}[!t]
	\centering
	\subfloat[\label{fig:compressor-performance-4-low}Low antenna correlation.]{\includegraphics[width=0.9\linewidth]{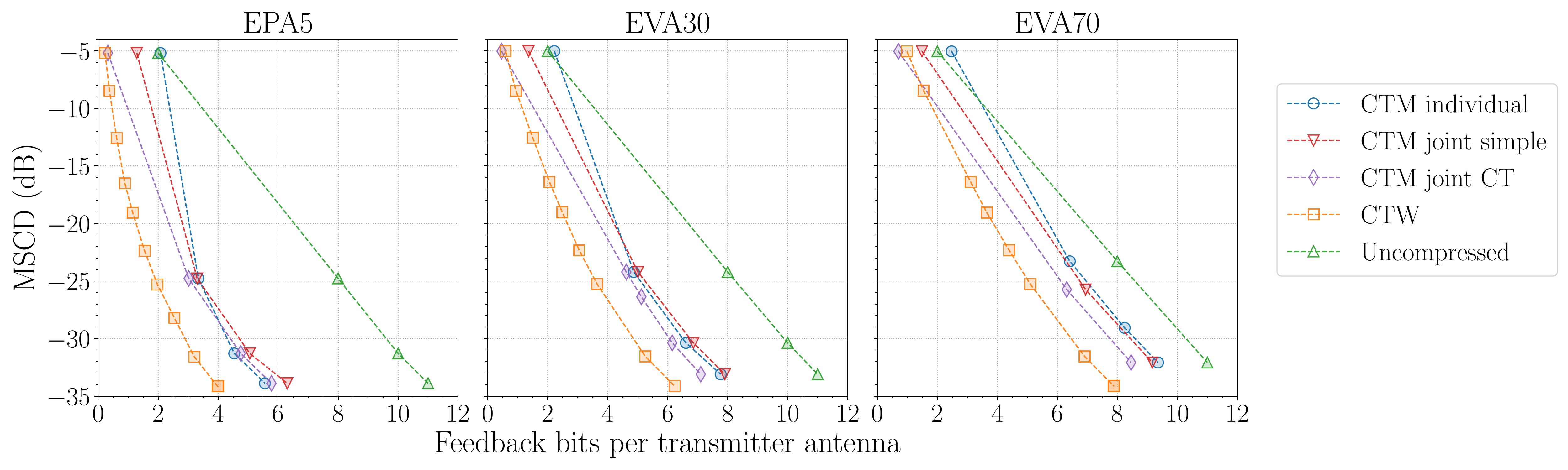}}\\
	\subfloat[\label{fig:compressor-performance-4-high}High antenna correlation.]{\includegraphics[width=0.9\linewidth]{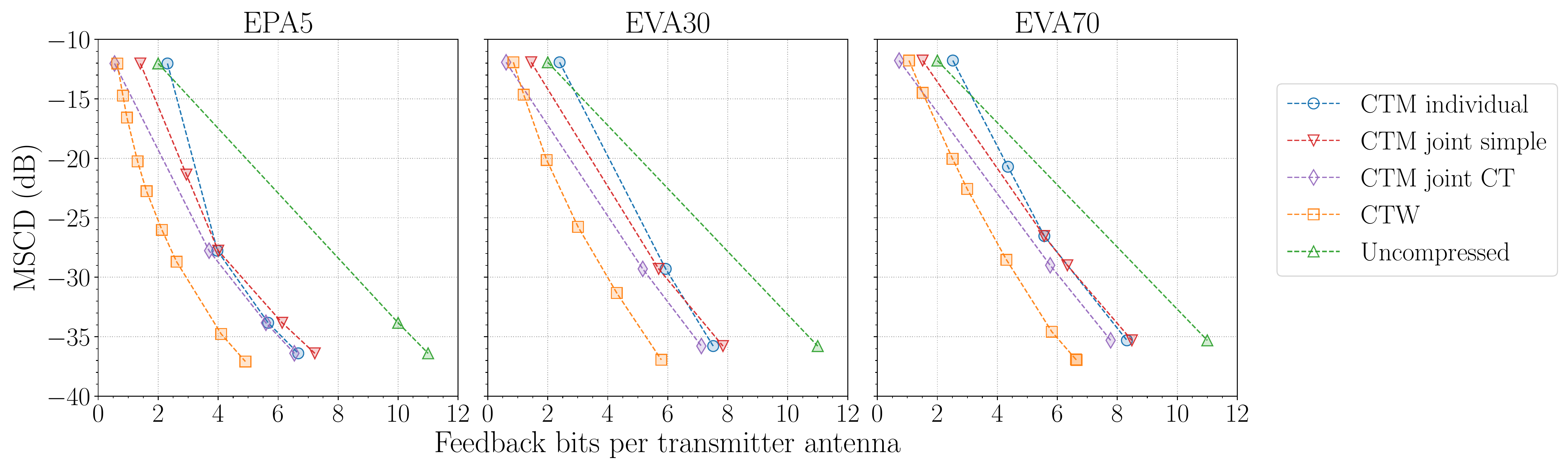}}\\
	\caption{MSCD distortion using $\beta$-law compander, for $\Nt=4$, $\Nr=4$.}
	\label{fig:compressor-performance-4}
\end{figure*}

\begin{figure*}[!t]
	\centering
	\subfloat[\label{fig:compressor-performance-16-low}Low antenna correlation.]{\includegraphics[width=0.9\linewidth]{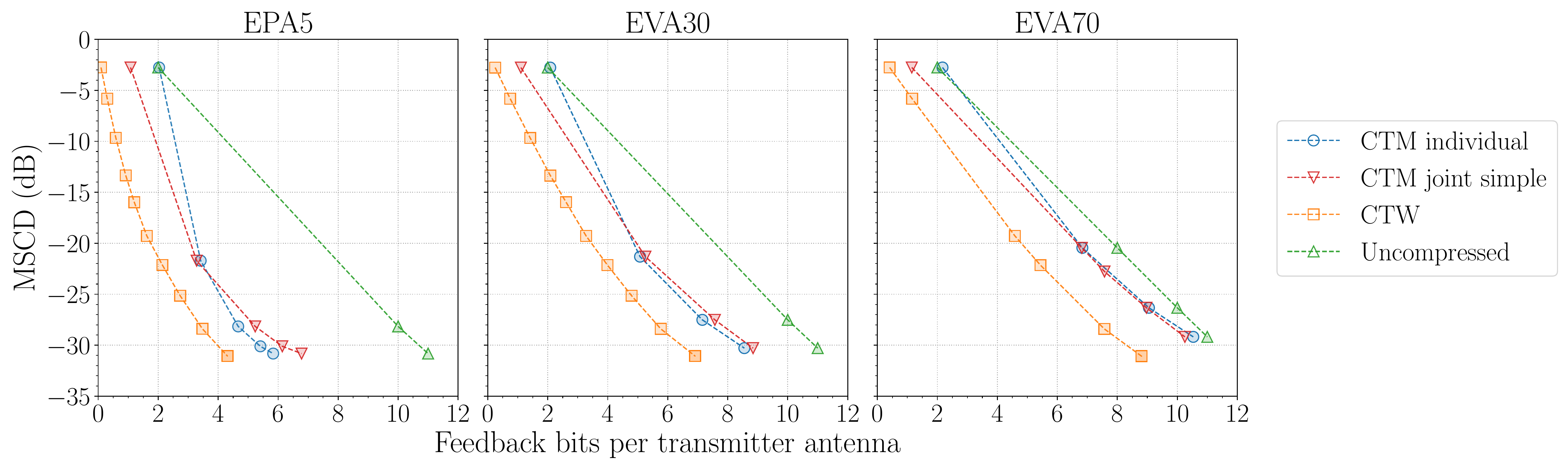}}\\
	\subfloat[\label{fig:compressor-performance-16-high}High antenna correlation.]{\includegraphics[width=0.9\linewidth]{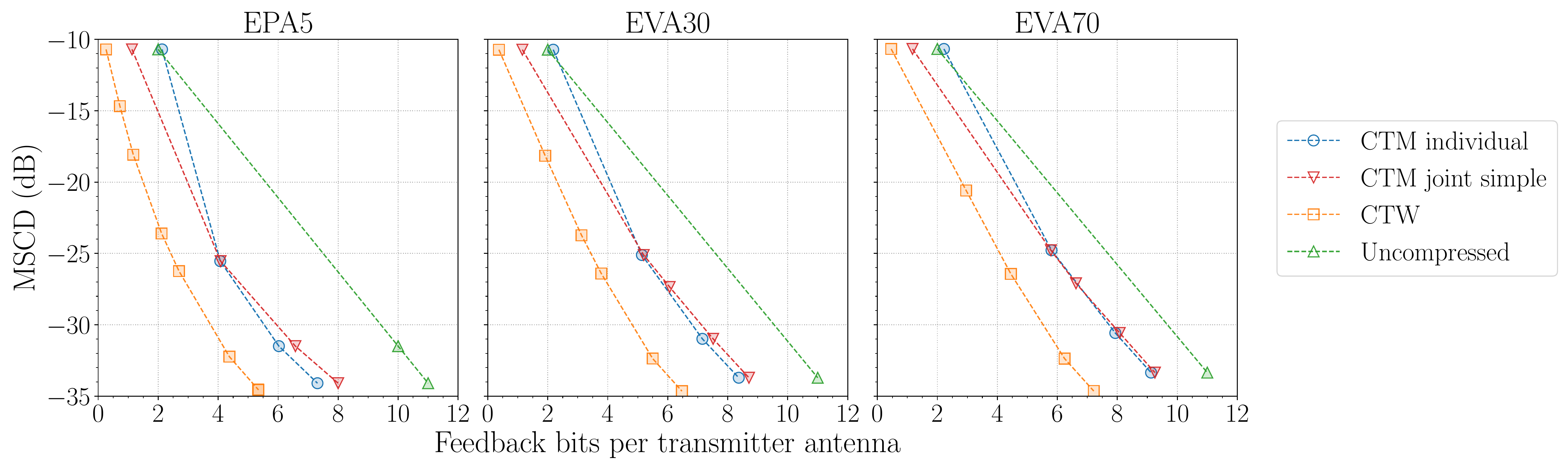}}
	\caption{MSCD distortion using $\beta$-law compander, for $\Nt=16$, $\Nr=8$.}
	\label{fig:compressor-performance-16}
\end{figure*}

Now, we fix the $\beta$-law compander as quantiser, and evaluate the performance of different compression schemes. In all cases, we assess the MSCD versus the average number of CSI bits per antenna, and plot the envelope formed by the best quantisation parameters~(different codebook sizes) over those that have been tested.

We compare different variations of the proposed context-tree two-level scheme---individual compression, joint compression with the simple strategy, and joint compression with context-tree~(CT) compression of the indicator sequence, cf.~Section~\ref{subsec:multiple-trees}---with uncompressed and ideal CTW combined with arithmetic coding. The ideal CTW case~\cite{willems-ctw} is simply evaluated with $\frac{1}{n} \left( \left\lceil -\log Q_n(x_1^n|x_{1-D}^0)\right\rceil + 1 \right)$. The performance of the different methods is studied for different mobilities and antenna correlations.

Fig.~\ref{fig:compressor-performance-4} shows the case $\Nt=\Nr=4$, for both low and high antenna correlation. Regarding the different CTM variations, we note that individual encoding of each sequence generally uses more bits, which is expected, as it does not exploit the spatial correlation between the antennas. Jointly encoding the sequences with the simple strategy can reduce the CSI bit rate, especially in low bit rate regime, and the CT compression of the indicator sequence can further compress the sequence.

Now, comparing the CTM with CT compression of the indicator sequence with uncompressed and CTW schemes, we see that the compression gains are significant and can reduce the CSI bit rate by up to a quarter in low correlation, and a half in high correlation, both in low rate regime. For EPA5, in the higher rate regime, the proposed CTM scheme can reduce the feedback in at least 4.5~bits and is 1.5~bits away from the CTW performance, approximately. For higher mobilities, the gains are more modest, due to the lower time correlation. Nevertheless, for EVA70, in the higher rate regime, the proposed CTM can save approximately 2.5~bits, and CTW, 4~bits, at least. Furthermore, for EVA30 and EVA70 in the extreme low rate regime, the proposed CTM slightly outperforms CTW, thanks to the auxiliary lower resolution quantiser.

Similarly, Fig.~\ref{fig:compressor-performance-16} shows the performances for the case $\Nt=16$, $\Nr=8$. In the lower rate regime, there is approximately a gain of 1~bit when using the CTM scheme with simple joint strategy, and an additional gain of up to 1~bit if CTW is used instead. In the high rate regime, the gains are more pronounced in the low mobility scenario. For instance, in EPA5, the CTM schemes can provide a saving of approximately 5~bits, which is less than 2~bits away from the CTW performance, in the low correlation case.

\subsection{Communication Rate}

\begin{figure*}[!t]
	\centering
	\subfloat[\label{fig:comm-rate-4}$\Nt=4$ and $\Nr=4$.]{\includegraphics[width=0.9\linewidth]{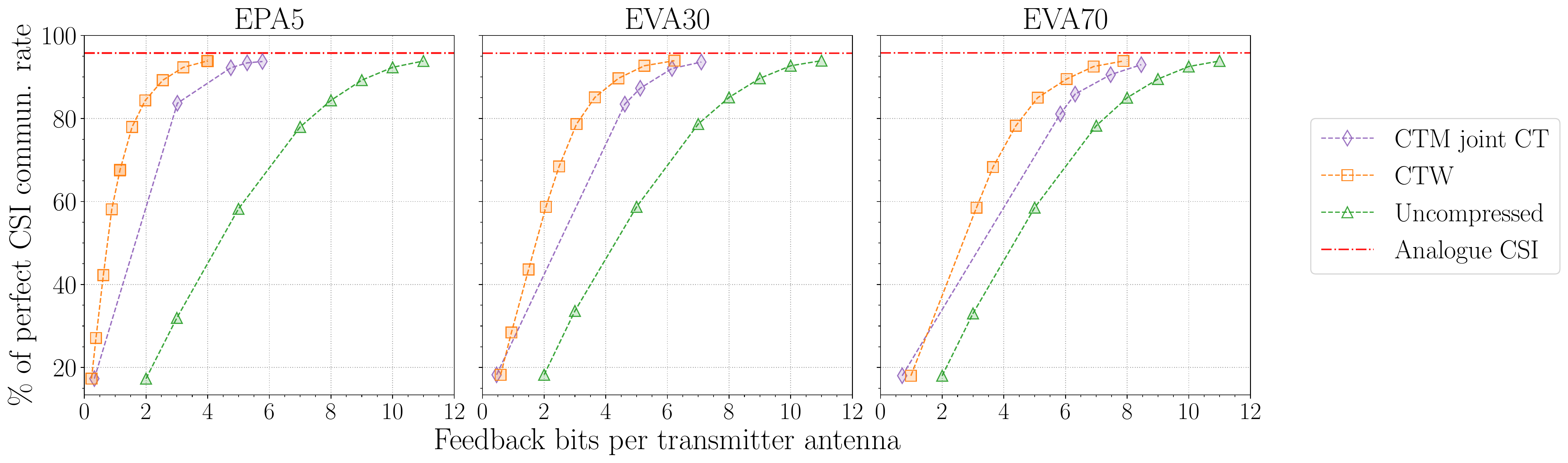}}\\
	\subfloat[\label{fig:comm-rate-16}$\Nt=16$ and $\Nr=8$.]{\includegraphics[width=0.9\linewidth]{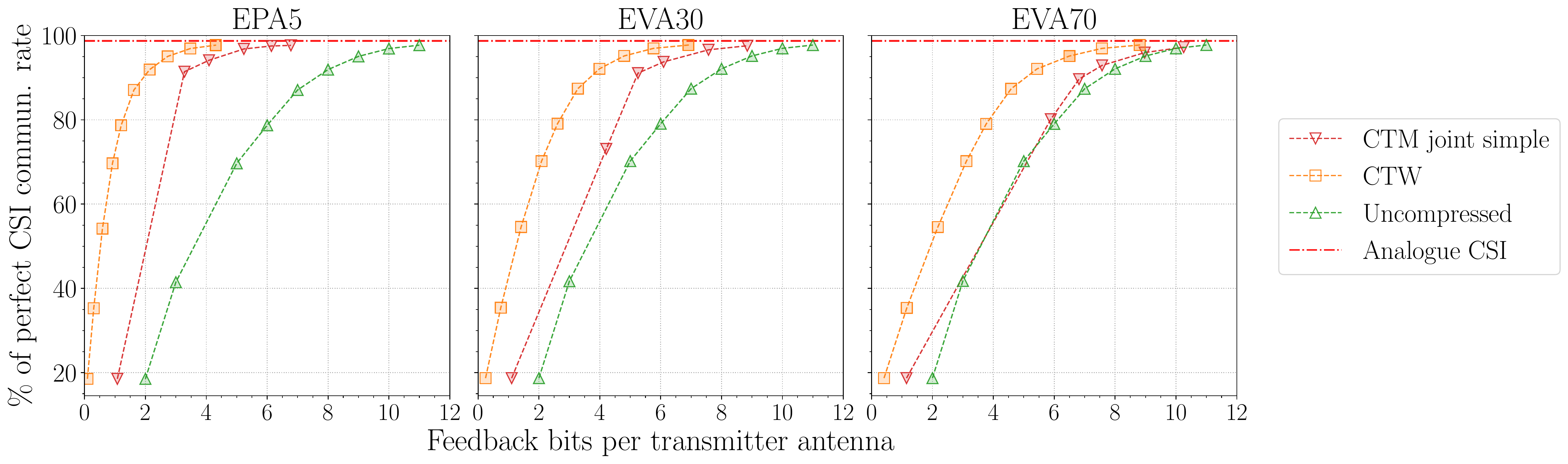}}
	\caption{Communication sum rate, using $\beta$-law compander.}
	\label{fig:comm-rate}
\end{figure*}

We also illustrate the gains in terms of the downlink communication sum rate with zero-forcing beamforming, evaluated approximately using the formula provided in~\cite[Eq. (20)]{caire}, for low antenna correlation. The results are normalised by the achievable rate when perfect~(i.e., noiseless) CSI is available to the BS, and are presented in Fig.~\ref{fig:comm-rate}, for different mobilities and number of antennas.

We emphasise that the communication rates converge much faster to the rate achieved by analogue CSI~(i.e., with no compression) when some of the proposed compression schemes is employed. For instance, for EPA5 and $\Nt=\Nr=4$, it takes the uncompressed scheme 11~feedback bits per antenna to achieve a communication rate close to the analogue upper-bound. Using the proposed CTM scheme, the same communication rate can be achieved with less than 6~bits, and, with ideal CTW, with 4~bits. In higher mobility, the gap between the two proposed scheme is smaller, while still presenting an advantage over not compressing at all. Similar gains are observed for $\Nt=16$, $\Nr=8$, even with the simple joint strategy.

\section{Conclusion} \label{sec:conclusion}
We have proposed a novel method for compressing CSI, combining lossy vector quantisation and lossless compression. The proposed vector quantiser is based on applying a data-adapted compander to the components of normalised vectors. The compression algorithm uses the estimated probability provided by the CTM model to encode a symbol, following a simple rule within a fixed number of levels. Simulations of LTE channels show the effectiveness of our approach in different scenarios.

More importantly, the proposed schemes have low complexity, can be implemented in an online fashion, and are modular. The context-tree-based compression scheme can be applied on any other quantisers, including those recently designed with neural networks, e.g.,~\cite{mashhadi}. Similarly, the proposed quantiser can be combined with any other lossless compression schemes.


\ifCLASSOPTIONcaptionsoff
  \newpage
\fi



\end{document}